\documentclass{emulateapj}
\usepackage{apjfonts}
\shorttitle{Infrared studies of GRB 050319}
\shortauthors{George et al.}
\begin{document}
\title{A late, infrared flash from the afterglow of  GRB 050319}

\author{Koshy George \altaffilmark{1}}
\altaffiltext{1}{Physical Research Laboratory, Navrangpura,  Ahmedabad, Gujarat 380009, India} 
%\email{koshyg@prl.res.in, orion@prl.res.in, chandra@prl.res.in, ashok@prl.res.in}

\author{Dipankar P. K. Banerjee \altaffilmark{1}}
\author{Thyagarajan Chandrasekhar \altaffilmark{1}}
\author{Nagarhalli M. Ashok \altaffilmark{1}}
\email{koshyg@prl.res.in, orion@prl.res.in, chandra@prl.res.in, ashok@prl.res.in}

\begin{abstract}
We report the detection of a bright, near-infrared flash from the afterglow of GRB 050319, 6.15 hours after the burst. The IR flash faded rapidly from J=13.12 mag. to J$>$15.5 mag. in about 4 minutes. There are no reported simultaneous observations at other wavelengths making it an unique event. We study  the implications of its late timing in the context of current theoretical models for GRB afterglows.
                                
\end{abstract}

\keywords{ gamma ray: bursts --- infrared: general}

\section{Introduction}

The observed properties of GRBs (gamma ray bursts) and their  afterglows are successfully 
explained by the  internal-external shocks model (Meszaros 2002; Piran 2004). In this 
frame-work the afterglow
 arises from energy dissipation of the relativistic flow from the GRB as it is slowed 
 down by the surrounding circumburst matter. The optical light curve of the afterglow 
 generally decays slowly following a  power law  (or a broken power law). However, in 
 some instances, a rapid flash  is observed to occur contemporaneously with  the GRB 
 proper.  Such a prompt flash has been observed in the optical rarely (Akerlof et 
 al. 1999; Li et al. 2003; Vestrand et al. 2005)  while the first   infrared flash was detected only  
 recently (Blake et al. 2005).  Such prompt  emission could 
be  caused by a reverse shock occurring from the interaction of  the outflow with the 
circumburst matter (Piran 2004) though internal shocks have also been invoked to consistently explain prompt emission (Vestrand et al. 2005; Blake et al. 2005). So far,  there has been no  detection  of a late flash from a GRB.
Here, we report the detection  of such a  late flash from 
GRB 050319 (George et al. 2005) which furthermore  is only the second IR flash to be ever detected from a GRB.  Since the origin of such a late flash is not easily explained by current theories of GRB afterglows, it becomes important to 
first establish the validity of the detection. We believe we establish this convincingly in Section 2 but even then - given the  novelty of the result -  present our observational detection with  due caution. 
In Section 3, we discuss the implications of the late timing of the present IR flash in the context of the fireball model.

\section{Observations and data reduction}

On 19 March 2005, 09:31:18.44 UT, GRB 050319 triggered the Burst Alert Telescope on board
 the Swift gamma-ray satellite (Krimm et al. 2005) . ROTSE-IIIb (Robotic Optical Transient
Search Experiment) responded to GRB 050319  in 9.2s, 27s after the burst and detected a 
16 mag. source  which faded down to  $\sim$ 18 magnitude about 940 seconds after the 
burst (Rykoff et al. 2005). We became aware of the GRB only 6 hours after the outburst through the  Gamma-Ray Burst Coordinate Network (GCN). Ongoing  observations of the  nova V574 Puppis were suspended, the GRB field was acquired and an initial 20s J band (1.25${\rm{\mu}}$m) image of the GRB field was immediately taken using the 1.2 meter Mt. Abu Infrared Telescope coupled with a 256x256 HgCdTe (NICMOS3) array near-IR imager/spectrograph. In this frame, which we designate as D1 in Table 1,  the IR transient (IRT) is significantly detected at 12.5$\sigma$ above the background level (Figure 1). We then took  10 more frames in this position each of 60s duration (designated frames S1 to S10). Subsequently we dithered the field to three adjacent positions, again taking 10 exposures of 60s each at each dithered position. Thus the J band observations spanned  $\sim$ 40 minutes. The log of  the observations is given in Table 1.The dithered frames were median-combined to generate the sky frame which was subtracted from the object image to give the sky subtracted image. 
Since we had only 2 field stars A and B in the IRT frame, we could not get
an astrometric position for the IRT - astrometry  needs 3 or more stars - directly on this frame. We first measured the pixel-offsets of the IRT with respect to  star A which is more centrally located than B (call these $\Delta$x and $\Delta$y).  Subsequently,
we took the first set of dithered frames - just after the IRT
detection - in which we had moved the field  South by
$\sim$ 60". In these dithered frames, four field stars appeared (two were 
A and B; the other two stars are discussed further in the following paragraph).  Although the IRT was absent in these frames
as it had faded, we could reliably allocate an apparent x,y
position to it since its offsets, $\Delta$x and $\Delta$y, with respect to
A were known. Thus we had four reference stars in the
frame permitting astrometry to be done. The RA and Dec. of the IRT, derived in this manner are $\alpha$ = 10:16:47.66 $\pm$ 0.02, $\delta$ = +43:32:55.6 $\pm$ 0.6  (J2000)  consistent with the  Swift UVOT co-ordinates of 
$\alpha$ = 10:16:47.76 $\pm$ 0.03, $\delta$ = +43:32:54.9 $\pm$ 0.5 (Boyd et al. 2005). The total systematic error in the derived IRT position, arising from the above approach, was estimated by applying the same techniques to the
V574 Pup nova field being studied prior to the IRT detection.
Here, we used the same number  of reference stars in similar x, y positions as in the IRT analysis, included  offsets for the dithered nova frames, and calculated the coordinates of 6 stars  around the IRT position.
We find the mean RA and Dec of these 6 stars to be  0.42" (1$\sigma$ = 0.23") east and 0.37" (1$\sigma$ = 0.49") south of their catalog values respectively. The star closest to the IRT position has RA $\&$ Dec offsets of
of 0.2"E and 0.18"S respectively. If we take into account this  systematic error, associated with a  fairly large 1$\sigma$ error, the derived IRT coordinates continue to be consistent with the UVOT position.  

\begin{table}
\caption{Log of observations. }
\begin{tabular}{ccccc}
\hline 
\hline\\
 Exposure Start      & Exposure     &  Frame        & $J$ Magni-\\
  (seconds)\tablenotemark{a}          & duration(sec)&  designation  & tude    \\
\hline  

  22151(= 6.15328 hours)    & 20      		&  D1           & 13.12 $\pm$ 0.08    \\
  22200              & 60           &  S1           & 14.55 $\pm$ 0.10    \\
  22261              & 60           &  S2	        & 14.81 $\pm$ 0.20    \\
  22322              & 60           &  S3	        & 14.79 $\pm$ 0.20    \\
  22383              & 60           &  S4	        & 15.29 $\pm$ 0.28    \\
  22444              & 60           &  S5	        & $>$15.5                \\

\hline
\hline\\
\tablenotetext{a} {Exp. start is relative to the Swift trigger of 2005 March at
   9.521789 UT }
\end{tabular} 
\end{table}

In the absence of a good sky flat, we have adopted a slightly different approach to 
correct for the effects of flat-fielding on the measured counts of the IRT and stars A $\&$ B by using the nova V574 Puppis field. The V574 Pup field is fairly crowded{\footnote {see www.prl.res.in/$\sim$chandra for an image plus other related material.}} and furthermore  images of it were obtained in 4 dithered positions. Thus in atleast one or more of these dithered images of the 
V574 Pup field, we could get a star (or stars) of this field to be sufficiently close  in 
array (x,y) coordinates to the (x,y) coordinates of the IRT or stars A $\&$ B.   
We assume, that the response of the array (i.e. its flat-field response)  will be 
reasonably similar over regions of the array separated by small amounts of $\sim$ 6 to 8 
pixels in x and y. Thus a comparison of the differential magnitudes of  the IRT and 
stars A $\&$ B with their closely-juxtaposed 2MASS counterparts in the V574 field, should 
reasonably account for flatfielding effects  and lead to correct $J$ magnitude estimates 
for the IRT and stars A and B. In effect, we are using not one but several stars of the 
nova field to act as calibrators. This  should ensure internal consistency and also reduce 
the scope for any major error in the derived magnitudes of the objects of interest. In 
this manner we obtain $J$ = 12.56 $\pm$ 0.03 mag. for star A and 13.98 $\pm$ 0.05 mag. for star B 
which are in reasonably good agreement with their 2MASS magnitudes of 12.466 $\pm$ 0.018 and 13.839 $\pm$  0.025 mags. respectively. In addition when the GRB field in Figure 1 was dithered northwards, two other 2MASS stars appear in the field below A and B.  These also are found to have 
closely-juxtaposed  counterparts in the V574 Pup field. Proceeding in a similar  manner 
as above, their $J$ band mags. are determined to be 10.85 $\pm$  0.03 and 11.24 $\pm$ 
0.03 respectively which again compare well with their 2MASS $J$ magnitudes of 10.827 $\pm$ 0.017 
and  11.279 $\pm$  0.017 respectively. Since the derived mags. of 4 field stars around the IRT match their 
2MASS magnitudes fairly well, we would thus believe that our derived  magnitudes for 
the IRT are  accurate.

Our observations were carried out under clear sky conditions. However the sky was bright 
in J band due to the presence of a ninth day  moon about 45$^0$ from the GRB position. This 
aspect has complicated the IR photometry. Our NICMOS3 detector has similar characteristics, 
pixel defects and other cosmetic artifacts as any other NICMOS3 detector used elsewhere. 
The detector is divided into 4 quadrants and unlike in an optical CCD, each quadrant is  
addressed separately during readout. The read noise and dark counts vary from quadrant to 
quadrant.  Strips and shading effects across the quadrants do exist  as seen in some of 
the frames in Figure 1.  But we note that similar artifacts are seen in other NICMOS 
detectors and that these could be understood in terms of settling down of  array after 
reset (Rieke at al. 1993 a,b; Meixner et al. 1999; e.g refer Figure 7 of Meixner et al. 1999) and 
a variability of the read-out noise  from pixel to pixel (refer Section 2.4 of Skinner et al. 1997 
and Figure 7 therein). A single column scan in the N-S direction of the array across the IRT position 
shows that the IRT signal clearly stands out, well above the background fluctuations due to shading  
patterns in the array.  These artifacts do not in fact affect the registration of a stellar image but 
can affect  the aperture  photometry. For e.g. in IRAF, while using APPHOT for aperture photometry,  the background 
counts to be subtracted from the stellar counts within a circular aperture centered on 
the star, is determined from the mean/median background counts in  an annulus positioned 
further away from the star. In case the annulus should fall on areas with contrasting 
background counts, the mean background count can be wrongly estimated. We have  taken 
care to avoid this error, especially in the case of the IRT  in frames 1 and 2 of 
Figure 1  -  in which an annulus is likely to sample varying backgrounds around the 
IRT because of shading - in the following way. Using the software IMPRO32, we have 
manually positioned a box aperture and obtained the counts around  the IRT or in the 
background as the case may be, while being extremely careful that the box  includes 
signal only from the appropriate regions.  Thus we are certain that we have determined the 
signal counts on the IRT and stars A and B with sufficient accuracy. In this context, we also point out that the (x,y) centroid of the IRT is measured to be at (126,133) pixels on the array.  Thus in the  direction of the array columns, in which the shading exists, the IRT is considerably off by 5 pixels off from the nearest quadrant edge (located at 128 pixels). This enables a 10 pixel square box-aperture to be positioned satisfactorily enough (around the IRT or the 
background) without including regions of varying intensity , which arise from shading, 
within the box.

Is it possible that the IRT registration is  an artifact of an unknown nature that 
arises by virtue of it either being (i) close to the array center or (ii) due to charge trapping? Both possibilities  appear unlikely. Regarding the first point, as mentioned before, the  (x,y) location of the IRT is not really coincident with the array centre at (128,128) but fairly well displaced from it. Also, from the survey of the 
literature on NICMOS3 array characteristics and behavior, we have not encountered   
mention of  any detector-related effect that causes a stellar-like artifact  to be created near the array center. Amplifier-glow, 
when the amplifiers are switched on during readout, is known to occur in NICMOS3 arrays 
but this is always seen at  the 4 corners of the array where the amplifiers are located, 
never from near the center. Regarding the second point of charge trapping, prior to the acquisition of the IRT afterglow a nearby field was imaged with an exposure of 10 seconds in which there was no object in the subsequent position of the IRT. Hence it can be discounted that the recorded image of the IRT is due to memory effects caused by charge trapping from previous images at the pixel position of the IRT. We have also carefully examined a faint bar-shaped artifact, seen in the frames of Figure 1, which is located in the top, left corner of the images and runs in the N-E to S-W direction.  We note that a single column scan across its brightest position  does not show significantly excessive  counts over the bar compared to the average background counts over the column. Further this artifact does not extend to the IRT position and we thus believe it does not  affect the detection. We have therefore carefully analysed all possible 
factors that can affect the images of Figure 1 and feel that we have established  that the IRT detection  is secure. 

\begin{figure}
\plotone{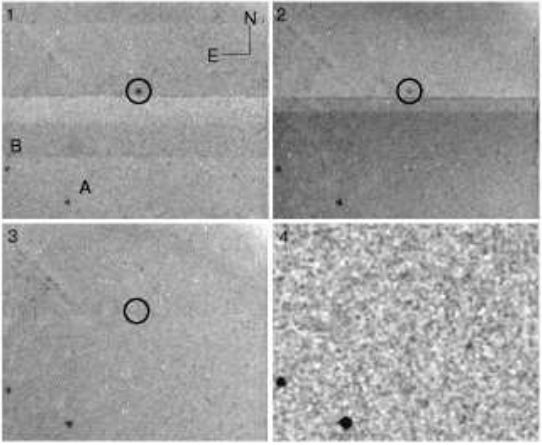}
\caption{1.25 ${\rm{\mu}}$m  J band images (3.2'x4') of the GRB 050319 field from the
1.2 meter Mt. Abu Infrared Telescope using the near-IR imager/spectrograph with a 
256x256 HgCdTe (NICMOS3) array under clear sky conditions. The IR afterglow (encircled) 
was detected in five images and was below the detection limit in the sixth. Here the 
first, second and sixth frames are presented  (marked 1, 2 and 3 here but  corresponding 
to entries D1, S1 and S5 of Table 1) to show the detection, fading and disappearance. 
The 2MASS field is shown in frame 4.  \label{fig1}}
\end{figure}

\begin{figure}
\plotone{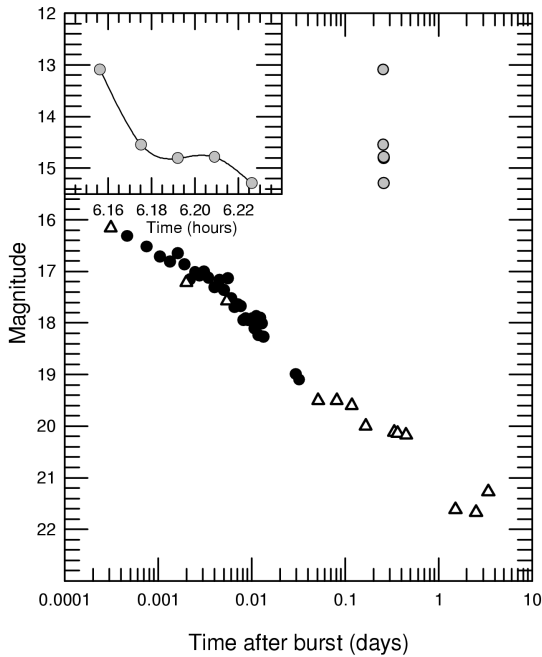}
\caption{ J band observations (filled grey circles) superposed on  the R band light 
curve of  GRB 050319 to accentuate the fast IR fading. The majority of the R band data (shown 
by filled circles) is  from Wozniak et al. 2005   while the rest (shown by triangles) 
are from  GCN circulars (Quimby et al. 2005; Yoshioka et al. 2005, Torii 2005; Sharapov et al. 2005a; Kiziloglu et al. 2005; Sharapov et al. 2005b ). The rapid fading of the IR flash may be seen 
here and also in the inset  showing  greater details. \label{fig2}}
\end{figure}

\begin{figure}
\plotone{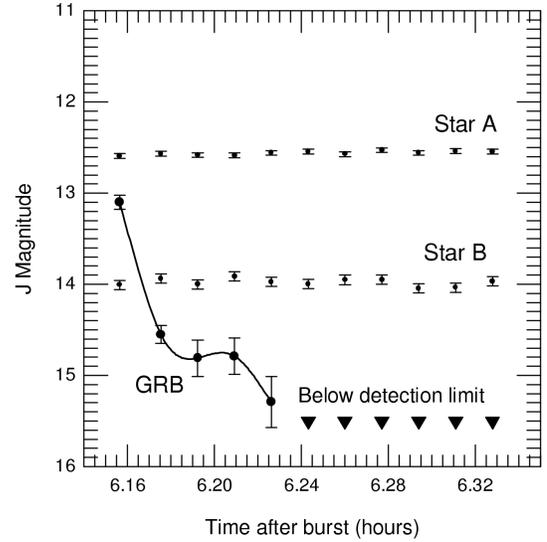}
\caption{ J band lightcurves of the IRT and of the 2 stars A and B in the field.  The  lightcurves of the field stars remain steady while the IRT fades rapidly, showing thereby  that the fading of IRT is genuine. \label{fig3}}
\end{figure}

\section{Results and Discussion}
A rapid dimming of the IRT is seen in the images of  Figure 1. This fast fading is also
depicted in the lightcurve shown in Figure 2. To clearly demonstrate  that the fading of 
the IRT is genuine we compare its lightcurve with those of stars A and B in Figure 3. The
lightcurves  in Figure 3 uses data from the first 11 frames. As can be seen, the lightcurves
of stars A and B remain stable around their mean magnitudes within $\pm$ 0.03 and $\pm$ 
0.05 mags respectively whereas the IRT fades. It  is difficult to say whether an optical 
equivalent of the IR flash occurred because there are no reported, concurrent observations - the closest R band optical data being 2.19 before (Yoshioka et al. 2005) and 1.84 hours after (Sharapov et al. 2005) our observations respectively. The closest reported V band data is from SWIFT,UVOT (Boyd et al. 2005), 18700 seconds after the burst, one hour prior to our observation. Thus a flare with $\sim$ 4 minute duration as we are recording here could easily have been missed in the above optical observations even if it had occured. A rapid brightness decline, on similar  time scales as reported here,  has been seen in the optical  in GRB 990123  ( Akerlof et al. 1999; a decline of 3 mags. in $\sim$ 110 
seconds; but note that GRB 990123 was detected in both the rising and declining phases) and
 in GRB 021211 ( Li et al. 2003; a 2 mag. drop in brightness in $\sim$ 300 seconds). In the IR the first detection 
of a flash was reported only very recently for GRB 041219 (Blake et al. 2005).  This IR flash, occurring 7.2 
minutes after the gamma-ray trigger, shows a source that brightens and fades rapidly in the
JHK bands - the total variability of  2.2 mags occurring in $\sim$ 90 seconds. GRB 041219 
however shows further complexities in its light curve with a rebrightening taking place 20 
minutes after the trigger. It is worth mentioning two other cases that are discussed 
(Piran 2004) in the 
context of optical flashes, more  because of the strong or early optical emission detected 
from them viz. GRB 021004 and GRB 030329.  GRB 030329 had a very bright 12 mag. afterglow
which faded by 0.2 mag in $\sim$ 860 seconds (Price at al. 2003) while GRB 021004 was 
detected at 15.45 mag.  
and showed a slow fading of $\sim$ 1.1 mag over 36 minutes (Fox et al. 2003). As may be seen,  the decline  in 
the afterglow brightness of these GRBs is  much slower than that seen in an optical 
flash proper.

At a redshift of z = 3.24 (Johan et al. 2005), GRB 050319 is one of the farthest cosmological GRBs.  
Assuming a Lambda cosmology with H${_{\rm 0}}$ = 71 km/s/Mpc,  $\Omega$${_{\rm M}}$ = 0.27 
and $\Omega$${_{\rm A}}$ = 0.73 the luminosity distance d${_{\rm L}}$ is found to be
 28.36 Gpc. An integrated fluence S = 8$\times$10${^{\rm -7}}$ erg/cm${^{\rm 2}}$ in the 15-350
KeV passband was measured by Swift (burst duration = 15s) for this GRB (Krimm et al. 2005). For such a value 
of S, the isotropic energy release for GRB 050319, calculated using E${_{\rm \gamma, iso}}$ = 
4$\pi$d${_{\rm L}}$$^{2}$S/(1+z), 
is found to be 1.8$\times$10${^{\rm 52}}$  ergs which is typical of the energy release  for GRBs.

The interpretation of the observed IR flash in context of the shocks model appears to
 be difficult.  We consider various mechanisms that can explain fluctuations in the 
 afterglow light curves viz.  reverse shocks, variations in the density profile of the 
 circumburst material and refreshed shocks.  The $\gamma$-ray emission in GRBs is believed to 
 originate from internal shocks when different 'shells' in a relativistic outflow from
the compact central source collide with each other. The afterglow is produced by the
interaction of this relativistic expanding flow   with the circumburst material.   A 
reverse shock, originating from this interaction, is predicted to occur contemporaneous
 with the prompt $\gamma$-ray emission and  give rise to a strong optical flash. Such a 
reverse shock is invoked to explain (Nakar $\&$ Piran 2005) the prompt optical flash seen for example in GRB
990123. However, in our case, the flash occurs 6 hours after the prompt $\gamma$-ray emission, 
well after the afterglow is visible, and is therefore extremely unlikely to be caused 
by a reverse shock. Further, generalized arguments - applicable to a reverse shock also -  indicate   that 
the duration of an observed variation (as in a flash or rebrightening)) should be  similar to the time elapsed after the burst (Piran 2004). Since this is not the case here,  a reverse shock is an unlikely cause for the observed IRT.  Refreshed shocks are caused  when slow shells in the ejecta 
catch up with the decelerating afterglow shock at later times causing a rebrightening of 
the afterglow light curve.  For such refreshed shocks too,  $\Delta$t is expected to be of the 
order of t (Kumar $\&$ Piran 2000). But there is a severe mismatch in the  time-scales $\Delta$t  and t here. 
Theoretical investigations have studied the effects of variations in the circumburst 
density  on afterglow lightcurves.  A fireball expanding into a wind with decreasing, 
outward density, as in the wind from  WR stars  which are considered  potential 
progenitors of a GRB in the collapsar model (Woosley 1993),  does not cause abrupt light-curve 
changes but rather leads to a steeper decline in the light curve (Chevalier $\&$ Li 1999)
vis-a-vis expected that in a constant density medium. More importantly, light-curve 
variations caused by  over/under-dense regions in the circumburst material have been 
simulated (Nakar $\&$ Piran 2003) for a 
variety of density profiles and  specifically applied to  GRB 021004 which showed 
a steep decay after a rebrightening at $\sim$ 4000s after outburst.  It is shown that the 
relatively fast decays (for e.g the decline of  $\sim$2.2 mags. in 10.5 hours seen in GRB 
021004 subsequent to its rebrightening)  cannot be reproduced by  any reasonable, 
realistic, spherically-symmetric density variation in the circumburst matter.  Thus 
in the present case, where the  variability time-scale of the IR flash is several 
orders smaller than  in GRB 021004, the difficulty in invoking density variations 
for causing the IR flash would become even more  magnified.  The most likely cause 
for the flash, as suggested for GRB 021004 also (Nakar $\&$ Piran 2003), could be the presence of  angular 
structures in the ejecta or within the external circumburst matter. Such smaller 
structures could reduce the angular smoothing time-scale and hence reduce the duration 
of a fluctuation. But detailed models are needed to confirm this. Alternatively  it 
needs to be assessed whether a dust echo around the progenitor can produce the 
characteristics of the observed flash. As has been pointed out recently, WC stars 
possess dust shells at typically 10$^{14}$-10$^{15}$ cms, and the echo of the initial GRB outburst 
from such a dust shell can produce variations in the GRB lightcurve on a timescale similar to what we observe here i.e. hours after the burst (Moran $\&$ Reichart 2005). In this context we also note optical observations of the afterglow of this GRB shows the emergence of an additional component about 10$^4$ seconds after the burst (Wozniak et. al 2005) which the authors have attributed to forward shock emission.

To summarise, we present evidence for an IR flash in GRB 050319 occurring $\sim$ 6.15 hours 
after the $\gamma$-ray emission which shows a rapid fading of 2.2 mags. in $\sim$ 4 minutes.  Since a late flash is unexpected, efforts have been made to demonstrate  convincingly  that the detection is beyond observational errors.  The present results could be suggestive of  a new aspect about GRB afterglows that is yet to be understood.
     
Research at the PRL is funded by the Department of Space, Government of India. We thank the referees for their comments that helped improve the paper.

%______________________________________________________________

%\clearpage

%\clearpage

%\clearpage

%\clearpage


\begin{thebibliography}{}


\bibitem[ref1]{r1}
Akerlof, C.  et al.
1999, Nature, 398, 400

\bibitem[ref2]{r2}
Blake, C.H.  et al. 
2005, Nature, 435, 181

\bibitem[ref3]{r3}
Boyd, P.  et al. 
2005, GCN circ, 3129


\bibitem[ref4]{r4}
Chevalier, R. A. $\&$ Li, Z.-Y. 
1999, \apj, 520, L29

\bibitem[ref5]{r5}
Fox, D.  et al. 
2003, Nature, 422, 284

\bibitem[ref6]{r6}
George, K., Jain, J. K., Ashok, N. M. $\&$ Chandrasekhar, T.
2005, GCN circ. 3126 

\bibitem[ref7]{r7} 
Johan, P.U. et al. 
2005, GCN circ. 3136 

\bibitem[ref16]{r16}
 Kiziloglu, U. et al. 
 2005, GCN circ. 3139 

\bibitem[ref17]{r17}
Krimm, H. et al. 
2005, GCN circ. 3117

\bibitem[ref20]{r20}
Kumar, P. $\&$ Piran, T. 
2000, \apj, 532, 286

\bibitem[ref21]{r21}
Li, W. et al. 
2003, \apj, 586, L9
 
\bibitem[ref27]{r27}
Meixner, M., Owl, R.Y. $\&$ Leach, R.W.  
1999, PASP, 111, 997

\bibitem[ref28]{r28}
Meszaros, P.
 2002, ARAA, 40, 137
 
\bibitem[ref29]{r29}
Moran, J. A. $\&$ Reichart, D. E. 
2005, \apj, 632, 438

\bibitem[ref30]{r30}
Nakar, E. $\&$ Piran, T. 
2005, \apj, 619, L147

\bibitem[ref31]{r31} 
Nakar, E. $\&$ Piran, T. 
2003, \apj, 598, 400

\bibitem[ref32]{r32}
 Piran, T.
 2004, Rev. Mod. Physics, 76, 1143

\bibitem[ref33]{r33}
Price, P.A. et al. 
2003, Nature, 423, 844

\bibitem[ref34]{r34}  
Quimby, R. M., Rykoff, E. S.,  Schaefer, B. E. $\&$ Mckay, T. 
2005, GCN circ. 3135 

\bibitem[ref35]{r35}
Rieke, M. J. et al.  
1993a,  Proc. SPIE, 1946, 179 

\bibitem[ref36]{r36} 
Rieke, M.J. et al. 
1993b, Proc. SPIE, 1946, 214 

\bibitem[ref37]{r37}
Rykoff, E., Schaefer, B., $\&$ Quimby, R.
2005, GCN circ. 3116 

\bibitem[ref38]{r38}
Sharapov, D. et al. 
2005a, GCN circ. 3124 

\bibitem[ref39]{r39}
Sharapov, D. et al. 
2005b, GCN circ. 3140 

\bibitem[ref48]{r48}
Skinner, C.J., Bergeron, L.E. $\&$ Daou.
1997 HST Calibration Workshop at STScI, eds. S. Casertano et al., 171 

\bibitem[ref49]{r49}
Torii, K. 
2005, GCN circ. 3121 

\bibitem[ref77]{r77}
Vestrand, W.T. et al. 
2005, Nature, 435, 178

\bibitem[ref50]{r50}
Wozniak, P. R. et al. 
2005, \apj, 627, L13


\bibitem[ref52]{r52}
Woosley, S. E. 
1993, \apj, 405, 273

\bibitem[ref53]{r53}
Yoshioka, T. et al. 
2005, GCN circ. 3120 


\end{thebibliography}
\end{document}